\documentclass[aps,prd,twocolumn,amsmath,amssymb,nofootinbib,superscriptaddress]{revtex4-1}
\usepackage{hyperref}
\usepackage{epsfig}
\usepackage{epstopdf}
\usepackage{upgreek}
\usepackage{graphicx}

\begin{document}

\newcommand{\bea}{\begin{eqnarray}}
\newcommand{\eea}{\end{eqnarray}}
\newcommand{\beq}{\begin{equation}}
\newcommand{\eeq}{\end{equation}}
\newcommand{\apjl}{ApJ}
\newcommand{\aj}{AJ}
\newcommand{\nar}{New Astron. Rev.}
\newcommand{\mnras}{MNRAS}
\newcommand{\apjs}{ApJS}
\newcommand{\aap}{A\&A}
\newcommand{\aaps}{A\&AS}
\newcommand{\jcap}{JCAP}
\newcommand{\Lam}{\ensuremath{\Lambda}}
\newcommand{\rmn}{\mathrm}

\title{Can a supervoid explain the Cold Spot?}

\author{Seshadri Nadathur}
\affiliation{Physics Department, University of Helsinki and Helsinki Institute of Physics, P.O. Box 64, FIN-00014, Helsinki, Finland}
\author{Mikko Lavinto}
\affiliation{Physics Department, University of Helsinki and Helsinki Institute of Physics, P.O. Box 64, FIN-00014, Helsinki, Finland}
\author{Shaun Hotchkiss}
\affiliation{Department of Physics and Astronomy, University of Sussex, Falmer, Brighton, BN1 9QH, UK}
\author{Syksy R\"as\"anen}
\affiliation{Physics Department, University of Helsinki and Helsinki Institute of Physics, P.O. Box 64, FIN-00014, Helsinki, Finland}

\date{\today}

\begin{abstract}
 The discovery of a void of size $\sim200\;h^{-1}$Mpc and average density contrast of $\sim-0.1$ aligned with the Cold Spot direction has been recently reported. It has been argued that, although the first-order integrated Sachs-Wolfe (ISW) effect of such a void on the CMB is small, the second-order Rees-Sciama (RS) contribution exceeds this by an order of magnitude and can entirely explain the observed Cold Spot temperature profile. In this paper we examine this surprising claim using both an exact calculation with the spherically symmetric Lema\^itre--Tolman--Bondi metric, and perturbation theory about a background Friedmann--Robertson--Walker (FRW) metric. We show that both approaches agree well with each other, and both show that the dominant temperature contribution of the postulated void is an unobservable dipole anisotropy. If this dipole is subtracted, we find that the remaining temperature anisotropy is dominated by the linear ISW signal, which is orders of magnitude larger than the second-order RS effect, and that the total magnitude is too small to explain the observed Cold Spot profile. We calculate the density and size of a void that would be required to explain the Cold Spot, and show that the probability of existence of such a void is essentially zero in $\Lambda$CDM. We identify the importance of \emph{a posteriori} selection effects in the identification of the Cold Spot, but argue that even after accounting for them, a supervoid explanation of the Cold Spot is always disfavoured relative to a random statistical fluctuation on the last scattering surface.
\end{abstract}

\maketitle


\section{Introduction}
\label{section:intro}

The existence of an anomalously cold region in the cosmic microwave background (CMB) at Galactic coordinates $(l,b)\sim(209^\circ,-57^\circ)$, known as the Cold Spot, was first reported in \cite{Vielva:2004} using data from the WMAP satellite analysed with a method based on Spherical Mexican Hat Wavelet (SMHW) wave functions. Several subsequent works \cite{Cruz:2005,Cruz:2006,Martinez-Gonzalez:2006,Cruz:2007a,Vielva:2010} have studied its statistical significance, morphology and non-Gaussianity using a variety of techniques. Analysis of data from the Planck satellite by the Planck team \cite{Planck:Isotropy} confirms the existence and location of the Cold Spot, and quantifies it anomalousness to be about $3\sigma$ in the standard $\Lambda$CDM model (although other authors differ on this question \cite{Zhang:2010,Bennett:2011}).

The existence of such a possibly anomalous structure has led to several proposed explanations for its existence, including a cosmic texture \cite{Cruz:2007b,Cruz:2008}, a large void along the line of sight \cite{InoueSilk:2006,InoueSilk:2007,Rudnick:2007kw}, a rare fluctuation on the last scattering surface, or a combination of these \cite{Inoue:2012}. In this paper we restrict our attention to the supervoid hypothesis.

Several theoretical studies (e.g., \cite{Martinez-Gonzalez:1990a,Martinez-Gonzalez:1990b,Panek:1992,Saez:1993,Arnau:1993,Fullana:1994,InoueSilk:2006,InoueSilk:2007,TomitaInoue:2008,Sakai:2008,Inoue:2010rp,Inoue:2012}) have been made of the secondary anisotropies on the CMB caused by large voids or overdensities,  through combinations of the integrated Sachs-Wolfe (ISW) effect \cite{Sachs:1967} (which we shall henceforth take to refer to the linear order effect) and its second-order counterpart, the Rees-Sciama (RS) effect \cite{Rees:1968}. However, these secondary anisotropies are much smaller than the primary fluctuations at last scattering, and these studies have invariably found that to explain the Cold Spot temperature decrement requires a supervoid of such large size and underdensity that its probability of existence in a $\Lambda$CDM universe is small. Of course, we must remain open to the possibility that such unusual structures could exist, and it is worthwhile to search for them -- especially since, if they do exist, they should produce other detectable signatures as well \cite{Das:2009,Kovetz:2013}.

Such structures have, however, not yet been found. A previous claimed detection of a supervoid at high redshift aligned with the Cold Spot direction \cite{Rudnick:2007kw} was later disputed \cite{Smith:2010,Bremer:2010,Granett:2010}. Claims have also been made for the existence of other large voids and superclusters leaving large ISW imprints on the CMB \cite{Granett:2008ju}, in contradiction to the $\Lambda$CDM expectation \cite{Hunt:2008wp,Nadathur:2011iu,Flender:2012wu}, but more recent studies using newer data do not find the same effect \cite{Cai:2013ik,Hotchkiss:2014}.

However, recently the detection of a new large void aligned with the Cold Spot direction has been reported based on analysis of the WISE-2MASS galaxy catalogue data \cite{Szapudi:2014}. This void is estimated to be centred at a redshift of $z\sim0.15$, have a size $\sim200\;h^{-1}$Mpc and a top-hat-averaged density contrast over this radius of $\bar\delta\sim-0.1$. Although the estimated spatial extent of the void is large, the density contrast is rather mild, and much smaller than the value that has previously been estimated to be required to explain the Cold Spot. In fact, the linear order estimate of the ISW temperature shift due to such a void is only $\Delta T\sim-20\;\upmu$K. Also, as we will argue in this paper, such a combination of size and density contrast is not even particularly unusual within the $\Lambda$CDM framework, as $\sim20$ such voids would be expected to exist within the local universe ($z<0.5$).

Nevertheless, it has been claimed that, if modelled using a spherically symmetric Lema\^itre-Tolman-Bondi (LTB) metric in a background $\Lambda$CDM model, this void can indeed provide an explanation for the full Cold Spot temperature decrement of $\sim-150\;\upmu$K \cite{Finelli:2014}. This is based on the argument that the LTB model can be described as a perturbation about a background homogeneous Friedmann--Robertson--Walker (FRW) metric such that, although the linear-order ISW effect is small, the second-order RS effect of the void is an order of magnitude larger and can account for the observed temperature profile around the Cold Spot direction. It is even claimed that a Bayesian model comparison strongly favours such a supervoid explanation of the Cold Spot over the alternative hypotheses of a cosmic texture or simply a statistical fluctuation on the last scattering surface.

Such an inversion of the magnitudes of the ISW and RS effects is however rather extraordinary, and appears to be incompatible both with previous supervoid estimates and with general perturbation theory expectations. In this paper we examine this claim in more detail. We calculate the temperature effect of the postulated void on the CMB both with an exact treatment of photon propagation in the LTB metric, and with second-order perturbation theory in $\Lambda$CDM. We will show that results from these two approaches agree well with each other. More importantly, both approaches show that the dominant temperature effect of a void such as that reported in \cite{Szapudi:2014} is in fact a \emph{dipole} anisotropy caused by our motion with respect to the void centre. 
The amplitude of this dipole contribution, while large compared to the other multipoles, is much smaller than other contributions to the kinetic dipole, so it would be unobservable in dipole-subtracted CMB maps. When the dipole contribution of the void is subtracted, we find that the remaining temperature signal is indeed about $-20\;\upmu$K, dominated by the linear-order ISW effect, and has an angular profile $\Delta T(\theta)$ that does not match the observation. Contrary to the claim in \cite{Finelli:2014}, we show that for reasonable void parameters the RS effect is always at least an order of magnitude smaller than the leading order ISW term.

We then turn to the issue of a fair comparison between theoretical models purporting the explain the observed $\Delta T(\theta)$ profile, and discuss the very important \emph{a posteriori} selection effect inherent in the identification of the Cold Spot using the SMHW technique. This selection effect must be correctly accounted for in assessing the probability of the null hypothesis that the observed profile is simply due to a statistical fluke. If this is done we find that in $\Lambda$CDM the probability of existence of a void that is large enough and deep enough to explain the Cold Spot temperature anomaly is always smaller than the probability that the original anomaly is simply due to a statistical fluctuation on the last scattering surface. This conclusion does not change even when combining the effect of the \emph{a posteriori} selection with the possible effect of any hypothesized void.
 
This means that in the absence of an actual detection of a void of such size and density as to directly challenge the otherwise successful $\Lambda$CDM model (or of a well-motivated theoretical reason for its existence), a fair analysis of Bayesian evidence will always disfavour any supervoid explanation of the Cold Spot, simply because the prior level of belief in its existence will be necessarily low. In this sense we argue that insofar as the Cold spot is anomalous, postulating a supervoid aligned with its direction cannot explain the anomaly.

The layout of the rest of the paper is as follows. Before discussing the details of the calculations, in Section ~\ref{section:heuristics} we first provide some simple heuristic arguments against the claim that the secondary effect of a void on the CMB could be large enough to explain the Cold Spot. In Section~\ref{section:voidCMB} we then explain the details of the calculation of this effect for the postulated void  profile and parameters, both using the LTB (Section~\ref{section:LTB}) and perturbation theory (Section~\ref{section:2PT}) approaches, and present our results. In Section ~\ref{section:Bayes} we then discuss the role of the selection effect and the appropriate approach to model comparison for competing explanations of the Cold Spot. Finally, we summarise and conclude our discussion in Section~\ref{section:discussion}. Some additional technical details and treatment of the special case $\Omega_\rmn{m}=1$ are presented in the appendices.


\section{Heuristic arguments against the supervoid}
\label{section:heuristics}

In \cite{Finelli:2014} the authors claim that the void reported from WISE-2MASS data \cite{Szapudi:2014} is capable of entirely explaining the observed CMB temperature profile around the Cold Spot, via the second-order RS effect. Before getting immersed in the technical details of the LTB and perturbation theory calculations in the next Section, it is worth first considering some simple intuitive arguments against such a claim.

The model of the void considered in \cite{Finelli:2014} has the density profile
\beq
\label{eq:deltadef}
\delta(r) = -\delta_0\left(1-\frac{2r^2}{3r_0^2}\right)e^{-\frac{r^2}{r_0^2}},
\eeq
where the best-fit parameters are claimed to be $\delta_0=0.25$ and $r_0=195\;h^{-1}$Mpc, while the void is centred at redshift $z_c=0.155$. This results in a top-hat-averaged density value for the void (at scale $r_0$) of $\bar\delta = -\delta_0/e\simeq-0.09$ \cite{Finelli:2014}. The maximum linear-order ISW temperature shift due to such a void from the linear-order ISW effect is just $\sim-20\;\upmu$K (on which point we agree with \cite{Finelli:2014}), whereas the Cold Spot shows $\Delta T\sim-150\;\mu$K at the centre.

The claim that such a void can explain the Cold Spot therefore relies on the argument that the second-order RS effect is an order of magnitude larger than the first-order ISW effect. But this seems completely at odds with the value of the density contrast ($\bar\delta$ or $\delta_0$), which seems to lie well within the linear regime. If the usual hierarchy of perturbation theory effects could be inverted for such mild density contrasts it would be hard to understand how linear theory predictions can successfully match any cosmological data. Nor can  the large size of the void provide an easy explanation for the relative importance of the RS term, since simple physical arguments indicate that both first and second-order terms have the same dependence on void size, $\Delta T_\rmn{ISW} \propto \delta r_0^3$ and $\Delta T_\rmn{RS} \propto \delta^2 r_0^3$.

The RS effect has been well studied in the literature. In an $\Omega_\rmn{m}=1$ universe for reasonable choices of the other cosmological parameters,  its amplitude is known to peak at $\Delta T\sim 0.1$-$1\;\upmu$K at multipoles $\ell\gtrsim100$ \cite{Seljak:1996}. Several studies have also considered the RS effect of model isolated voids or overdense structures in an $\Omega_\rmn{m}=1$ background \cite{Martinez-Gonzalez:1990a,Martinez-Gonzalez:1990b,Panek:1992,Saez:1993,Arnau:1993,Fullana:1994}: typically the structures modelled here have $\bar\delta\sim-1$ and size $\sim10$-$100\;h^{-1}$Mpc, but produce effects of at most $\Delta T\sim1$-$10\;\upmu$K. In suggesting a supervoid as an explanation for the Cold Spot, \cite{InoueSilk:2006} used a model void with $\bar\delta\sim-0.3$ at a radius $>200\;h^{-1}$Mpc in a $\Lambda=0$ universe. For $\Lambda\neq0$, it has since been shown \cite{InoueSilk:2007,TomitaInoue:2008} both that the RS effect of such model voids is subdominant to the linear-order effect, and that that the magnitude of the RS effect itself decreases as the value of $\Omega_\rmn{m}$ decreases from 1.

Perhaps more importantly, results from $N$-body simulations  \cite{Cai:2010hx} show that, within $\Lambda$CDM, on degree angular scales and at redshifts $z\lesssim1$ the RS temperature  is always negligible compared to the first-order ISW effect, and the maximum amplitude of the ISW+RS effect is an order of magnitude smaller than the $\Delta T\sim-150\;\upmu$K of the Cold Spot. This result was obtained for a (relatively) small simulation volume of $1\; h^{-3}$Gpc$^3$, but using a  simulation volume of $216\;h^{-3}$Gpc$^3$ it has since also been shown that the cumulative amplitude of the ISW signal of all structures out to redshift $z=1.4$ is at most $\sim50\;\upmu$K over the whole sky \cite{Watson:2013cxa}.

These results already suggest that any structure that \emph{could} explain the Cold Spot temperature must be an extremely rare fluctuation in $\Lambda$CDM. The void reported in \cite{Szapudi:2014} is however not particularly rare. Those authors estimate it to be a $3-5\sigma$ fluctuation; however this refers only to the value of $\bar\delta$ in units of the rms fluctuation of the density field $\sigma$ on the same $195\;h^{-1}$Mpc scale, and not to the probability of finding such structures in a large universe. Indeed, $\sim5$ voids with radii $150\lesssim R\lesssim300\;h^{-1}$Mpc and central underdensity $\delta_0<-0.7$ have already been found at redshifts $z\lesssim0.4$ in luminous red galaxy (LRG) catalogues from the SDSS  \cite{Nadathur:2014a}. (This density contrast value refers to the LRG density field, but taking a linear bias value $b\sim2$ as is appropriate for LRGs, one would still find that these voids are both deeper and larger than the ``supervoid'' reported in \cite{Szapudi:2014}.\footnote{Note that the density profiles of such voids \cite{Nadathur:2014b} are also similar to that postulated by \cite{Finelli:2014}.}) Yet they lie in the Northern Galactic hemisphere, where no Cold Spot-like CMB structures are observed. Indeed $N$-body simulations suggest both that in a full-sky survey one would expect to see $\sim20$ such voids, and that their ISW imprints on the CMB should be small \cite{Hotchkiss:2014}. If the claimed magnitude of the RS effect of such voids were correct, one would then expect to see several Cold Spot-like structures on the sky rather than only the one.  Indeed it has been argued that there are fewer hot and cold spots in the CMB than expected in $\Lambda$CDM \cite{Ayaita:2010}, an effect related to the relative absence of power on large angular scales.

As we will show in the next Section, the resolution of this puzzle is simple: correct calculation of the gravitational effects of the reported void shows both that the second-order RS contribution is subdominant as expected and that the total temperature anisotropy produced by the void is insufficient to account for the Cold Spot.


\section{The impact of a void on the CMB}
\label{section:voidCMB}

We will approach the calculation of the temperature anisotropies due to the void in two separate ways: one using an exact general relativistic calculation in the spherically symmetric LTB metric, and the other by treating the void as a spherically symmetric perturbation about a background FRW metric. 

Some previous works (e.g. \cite{VanAcoleyen:2008,Biswas:2008}) have established a procedure for mapping LTB metric solutions to the equivalent perturbation in FRW, and have studied the conditions under which the implied approximations are valid. However, all of these mappings have been studied for the pure dust ($\Lambda=0$) LTB model, in which case closed-form parameteric solutions to the Einstein equations are known \cite{Celerier:2000}. In the case of the LTB model with non-zero $\Lambda$, the equations are more complex, and we are not aware of any rigorous study of the conditions under which these treatments are equivalent.

The authors of \cite{Finelli:2014} model their void using an LTB metric and claim (but do not prove) that this is equivalent to a particular form of the potential fluctuation $\Phi$ about an FRW background. We will also not directly examine the approximations under which these two approaches are equivalent. Instead we will choose the spatial form of $\Phi$ to match, at linear order, the density profile of the LTB void. The time dependence of $\Phi$ will then be set by the cosmological model, which we take to be flat $\Lambda$CDM with fiducial parameters $\Omega_\rmn{m}=0.27$, $\Omega_\Lambda=0.73$. We then calculate the temperature anisotropies by the two different methods and show that -- for the void parameter values given in \cite{Finelli:2014} -- they give similar results.

\subsection{Exact LTB model}
\label{section:LTB}

The LTB metric has the spherically symmetric line element
\beq 
\label{eq:LTBmetric}
ds^2=-dt^2+\frac{R_{,r}^2(r,t)}{1+2E(r)} dr^2 + R^2(r,t) d\Omega^2,
\eeq 
where $d\Omega^2 = d\theta^2+\sin(\theta)^2d\phi^2$ and $R_{,r}\equiv dR/dr$. The function $E(r)$ represents a position dependent curvature, and the line element reduces to FRW in the special case $R(r,t)=a(t)r$ and $E(r)=-\frac{1}{2}kr^2$. For dust and a cosmological constant $\Lambda$, the Einstein equations reduce to
\beq
\label{eq:LTB_eom}
R_{,t}^2 = 2E(r) + \frac{2M(r)}{R} + \frac{1}{3} \Lambda R^2,
\eeq
where $M(r)>0$ is a free function related to the matter density by
\beq
\label{eq:rhoandM}
\rho(r,t) = \frac{M_{,r}}{4 \pi G R^2 R_{,r}}.
\eeq
The solution to the equation of motion \eqref{eq:LTB_eom} can be written as an integral equation
\beq
\label{eq:integsoln}
t-t_B(r) = \int_0^{R(t,r)} \frac{dA}{\sqrt{\frac{2M(r)}{A}+2E(r)+\frac{1}{3}\Lambda A^2}},
\eeq
where $t_B(r)$ is another free function known as the bang time.

The LTB model can thus be specified by the choice of three time-independent functions $E(r)$, $t_B(r)$ and $M(r)$, one of which corresponds to a gauge degree of freedom in redefining the radial coordinate. However, a spatially varying bang time $t_B(r)$ corresponds to a decaying mode \cite{Zibin:2008} which is incompatible with the standard cosmological picture of a universe that was very close to homogeneous at early times and is contradicted by observations of the CMB. Therefore the bang time is homogeneous in any realistic model, and can, without loss of generality, be set to $t_B=0$. 

We choose the curvature function $E(r)=E_0r^2\exp(-r^2/r_0^2)$ to match that in \cite{Finelli:2014}, which ensures that the LTB model asymptotically approaches a background FRW metric at large $r$. Those authors do not specify their choice of gauge, which is usually set by either $M(r)\propto r^3$ or $R(r,t_0)=r$ at the current time $t_0$. However, they refer to earlier work \cite{Garcia-Bellido:2008a} which uses the second gauge choice, and they provide an expression for the density contrast of the void as in eq.~\eqref{eq:deltadef}, which we take to define the density at $t_0$. We therefore choose the gauge by setting
\beq
\label{eq:M1}
M(r) = \frac{4 \pi G r^3}{3} \bar\rho(t_0) \left[1+ \delta(r)\right]\,
\eeq
where $\bar\rho(t)$ is the background FRW density. From eq.~\eqref{eq:rhoandM}, this is equivalent to choosing $R(r,t_0)=r$ if $\delta(r,t_0)$ matches eq.~\eqref{eq:deltadef} today.\footnote{It is possible that this condition and eq.~\eqref{eq:integsoln} cannot be simultaneously satisfied with a homogeneous bang time. Indeed we find that this is generally the case. For the fiducial parameter values we consider, the deviation of the true density profile from eq.~\eqref{eq:deltadef} is very small and has negligible effect. However, this deviation increases at large $\delta_0$ and may be partially responsible for the differences in $\Delta T$ seen in Figure~\ref{fig:d0scaling}.} The constant $E_0$ can then be chosen to match the required value of $\delta_0$.

Given these choices of the free functions of the model, we solve the integral equation in eq.~\eqref{eq:integsoln}. For $E(r) \neq 0$ and $\Lambda \neq 0$, it is not possible to express $R(r,t)$ in terms of elementary functions, and one must solve an elliptic integral numerically. We do this using Carlson's elliptic integrals \cite{Carlson:1995}, following the method outlined in \cite{Valkenburg:2012}.

Having thus obtained $R(r,t)$ and its derivatives, we may then write the equations for a null geodesic $k^\mu = \frac{dx^\mu}{d\lambda}$ as follows:
\bea
\frac{dt}{d\lambda} &&= 1+z, \\
\frac{d\theta}{d\lambda} &&= 0, \\
\frac{d\phi}{d\lambda} &&= \frac{c_\phi}{R^2}, \\
\frac{dz}{d\lambda} &&= -\frac{R_{,tr}}{R_{,r}} (1+z)^2+\frac{c_\phi^2}{R^2}\left(\frac{R_{,tr}}{R_{,r}} - \frac{R_{,t}}{R}\right),  \\
\frac{d^2 r}{d\lambda^2} &&= -2\frac{R_{,tr}}{R_{,r}}(1+z)\frac{dr}{d\lambda} - \left( (1+2E) \frac{R_{,rr}}{R} - E_{,r} \right) \left(\frac{dr}{d\lambda}\right)^2 \nonumber \\
 &&\;\;\;+ (1+2E)\frac{c_\phi^2}{R^3 R_{,r}},
\eea
where $z$ is the redshift measured by an observer in the dust rest frame, $1+z = k^{\mu} \delta^0_{\mu}$, the normalization is set so that $k^0=1$ at the observer, the angular coordinates are fixed by the choice $\theta = \pi/2$ and $c_\phi$ is an integration constant related to the impact parameter. The observer position is set such that the void centre lies at a distance corresponding to the comoving distance to redshift $z_c$ in the background FRW model.

The initial condition for the radial component of the tangent vector can be solved from the null condition $k_\mu k^\mu = 0$,
\beq
\frac{dr}{d\lambda} = \pm \frac{1}{R_{,r}} \sqrt{(1+z)^2 - \frac{c_\phi^2}{R^2}}.
\eeq
If the radial coordinate of the observer is $r_i$ and the angle made by the photon path with respect to the center of the void is $\alpha$, then $c_\phi = R(t_0, r_i) \sin(\alpha)$. We solve the geodesic equations backwards in time from today to some initial time at which the beam is well outside the void. In practice this was chosen to be $t_i = 0.1 t_0$. Since the temperature scales as $T \propto 1+z$, the temperature shift along this direction is then obtained from 
\beq
\frac{\Delta T}{T} = 1 - \left(1+z(\mathbf{\hat{n}},t_i)\right) a(t_i), 
\eeq
where $a(t)$ is the background FRW scale factor and $\hat{n}$ is the direction on the sky.

\begin{figure}
\includegraphics[width=80mm]{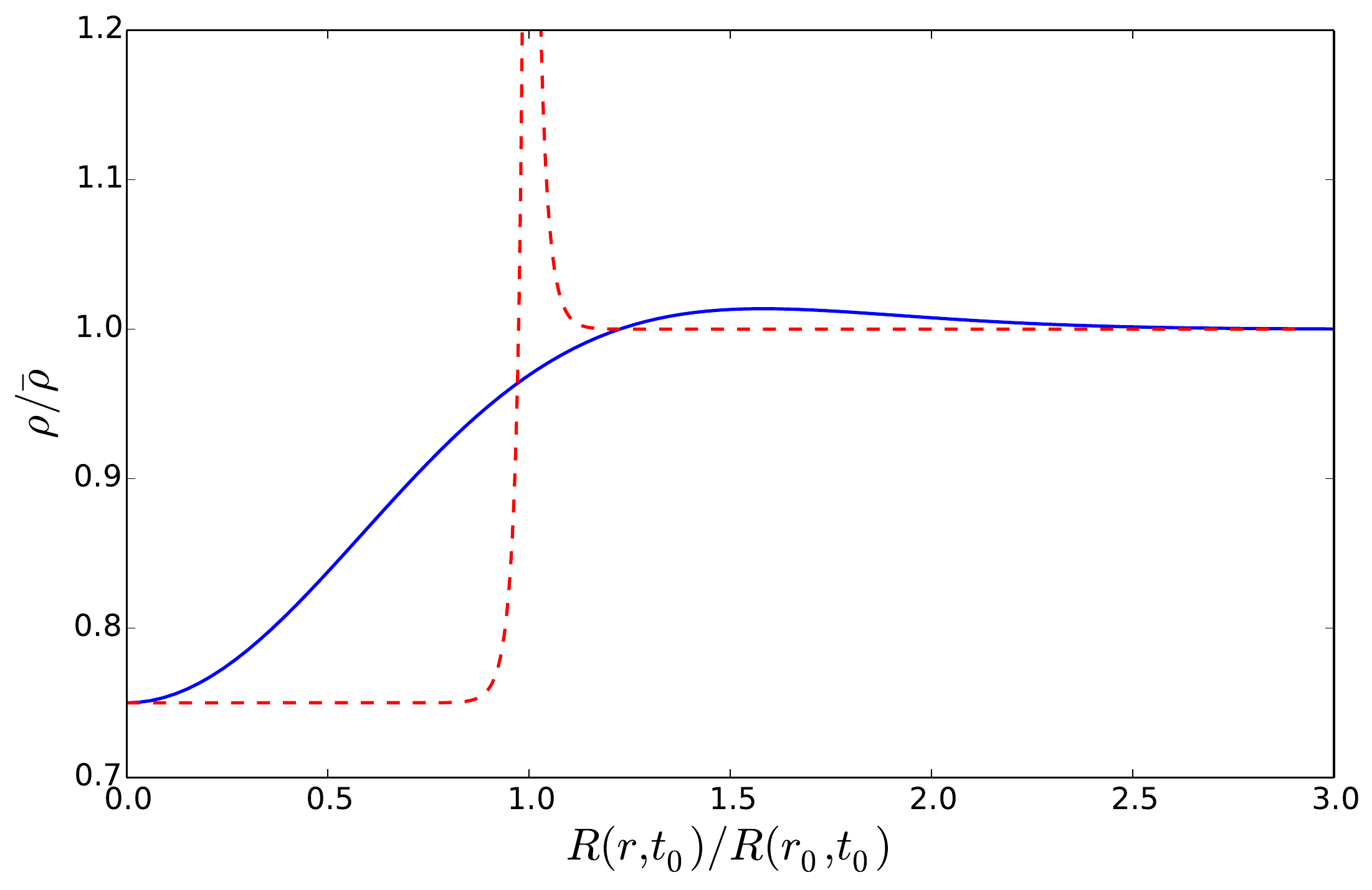}
\caption{Density profiles for the two different void models considered, at the current time $t_0$. The blue (solid) curve is for the fiducial void density profile of eq.~\eqref{eq:deltadef}. The red (dashed) curve is for an alternative model with a compensated top-hat profile, referred to as cLTB. The x-axis shows distance from the centre of the void in units of the void size as the gauge-independent ratio $R(r,t_0)/R(r_0,t_0)$ since the gauge choice is different for the two models.} 
\label{figure:densities}
\end{figure}

In addition to the model described above which is chosen to match that of \cite{Finelli:2014}, we also calculate the temperature anisotropies for a second choice of $M(r)$ and $E(r)$ to see how the results depend on these choices. The second model is specified by
\begin{align}
M(r) &= \frac{4 \pi G r^3}{3} \bar\rho(t_0), \\
E(r) &= E_0 r^2 \left[1 -\tanh \left(35 r/r_b - 20 \right)\right], \\
t_B(r) &= 0.
\end{align}
In contrast to the previous model, this describes a compensated top-hat-like density profile, with compensation radius $r_b$. We shall refer to this model as cLTB, for ``compensated LTB". Figure~\ref{figure:densities} shows the density profiles for the two cases. For the cLTB model we have fixed the value of $E_0$ to obtain the same density contrast $\delta_0=0.25$ in the centre, and the value of $r_b$ by requiring that the top-hat-averaged density to radius $r_0$ be equal to that for the density profile of eq.~\eqref{eq:deltadef}, i.e.,
\beq 
\label{eq:rbdef}
\frac{3}{R(t_0,r_0)^3} \int_0^{r_0} dr R^2(t_0,r) R_{,r}(t_0,r) \delta_\rmn{cLTB}(r) = -\frac{\delta_0}{e}.
\eeq
This gives $r_b = 1.72 r_0$.

\subsection{Perturbation theory model}
\label{section:2PT}

We consider perturbations about a flat Robertson-Walker space-time, for which the line element is
\beq
\label{eq:lineelement}
ds^2\simeq a^2(\eta)\left( \bar{g}_{\mu\nu}^{(0)}+\bar{g}_{\mu\nu}^{(1)}+\frac{1}{2}\bar{g}_{\mu\nu}^{(2)} \right)dx^\mu dx^\nu,
\eeq
where $\eta$ is the conformal time ($d\eta=dt/a(t)$), and $a(t)$ is the scale factor of the universe. The general metric can be written as
\beq
\bar{g}_{00} = -\left(1+2\psi^{(1)}+\psi^{(2)}+\ldots \right),
\eeq
\beq
\bar{g}_{0i} = z_i^{(1)}+\frac{1}{2}z_i^{(2)}+\ldots,
\eeq
\beq
\bar{g}_{ij} = \left(1-2\phi^{(1)}-\phi^{(2)}\right)\delta_{ij}+\chi_{ij}^{(1)}+\frac{1}{2}\chi_{ij}^{(2)}+\ldots,
\eeq
where the functions $\psi^{(r)}$, $z_i^{(r)}$, $\phi^{(r)}$ and $\chi_{ij}^{(r)}$ represent the $r$-th order metric perturbations.

A general gauge-invariant treatment of the CMB anisotropies due to such metric fluctuations up to second order in perturbation theory was presented in \cite{Mollerach:1997}, following \cite{Pyne:1996}, and we quote here the results of relevance to this work. The first-order temperature anisotropy $\delta T^{(1)}\equiv\Delta T^{(1)}/T$ is given by
\beq
\label{eq:1stdT}
\delta T^{(1)}= -\psi_\mathcal{O}^{(1)}+v_\mathcal{O}^{(1)i}e_i-I_1,
\eeq
where $v^{(1)i}$ is the first-order velocity perturbation, $e_i$ denote the basis vectors, subscript $\mathcal{O}$ refers to the observer's location, and we have suppressed all terms that depend only on variables at the last scattering surface. Thus the first term in eq.~\eqref{eq:1stdT} is a monopole, the second represents a dipole due to the observer's motion, and the first-order ISW contribution is given by
\beq
\label{eq:I1}
\delta T_\rmn{ISW} = -I_1 = \int_{\eta_\mathcal{E}}^{\eta_\mathcal{O}} d\eta \left(\psi^{(1)\prime}+\phi^{(1)\prime}\right)
\eeq
where the $^\prime$ denotes the derivative with respect to conformal time. For brevity we have suppressed vector and tensor contributions to this integral, since we will consider only scalar perturbations at linear order.

The expression for the second-order anisotropy is rather more tedious, but we will require only the part corresponding to the second-order Rees-Sciama term,
\beq
\label{eq:I2}
\delta T_\rmn{RS} = \frac{1}{2}\int_{\eta_\mathcal{E}}^{\eta_\mathcal{O}} d\eta\left( \psi^{(2)\prime}+\phi^{(2)\prime}\right).
\eeq
Here again we have pre-emptively dropped the second-order vector and tensor perturbation terms -- although these are not zero they will be negligibly small for the sub-horizon modes of interest to us \cite{Mollerach:1997,TomitaInoue:2008}.

In order to evaluate the first- and second-order contributions of eqs. \eqref{eq:I1} and \eqref{eq:I2}, we require the solutions for the perturbations $\psi$ and $\phi$. We choose to work in the Poisson gauge \cite{Bertschinger:1996}, the second-order generalisation of the longitudinal gauge. The Poisson gauge solutions of the Einstein equations up to second-order were presented in \cite{Mollerach:1997} for the special case a dust universe ($\Lambda=0$), and have been used in some previous works modelling the Rees-Sciama effect of voids \cite{Biswas:2008,Masina:2009a}. The general solutions for a flat universe with non-zero $\Lambda$ were first derived in \cite{Tomita:2005a} (see also \cite{TomitaInoue:2008}). 

For this case, the relevant pieces of the first-order solutions can be written in terms of two functions $P(\eta)$ and $F(\mathbf{x})$ as
\beq
\label{eq:phi1}
\psi^{(1)}=\phi^{(1)} = -\frac{1}{2}\left(1-\frac{a^\prime}{a}P^\prime\right)F,
\eeq
and
\beq
\label{eq:v1}
v^{(1)i}=\frac{1}{2}P^\prime F_{,i},
\eeq
where the growing mode solution for $P$,
\beq
\label{eq:P}
P = \int_0^\eta d\eta^\prime a^{-2}(\eta^\prime)  \int_0^{\eta^\prime} d\eta^{\prime\prime} a^2(\eta^{\prime\prime}),
\eeq
is determined by the cosmological model (i.e., the value of $\Omega_\Lambda$) alone.

The second-order perturbations in the same Poisson gauge can be written as \cite{Tomita:2005a,TomitaInoue:2008}
\beq
\label{eq:phi2}
\phi^{(2)}=\psi^{(2)}=\zeta_1 F_{,i}F_{,i}+\frac{9}{2}\zeta_2 \Psi_0,
\eeq
where 
\beq
\label{eq:Psi_0}
 \nabla^2 \Psi_0= F_{,ij}F_{,ij} - \left( \nabla^2 F\right)^2,
\eeq
\beq
\label{eq:zeta_1}
\zeta_1 = \frac{1}{4}P\left(1-\frac{a^\prime}{a}P^\prime\right),
\eeq
and
\beq
\label{eq:zeta_2}
\zeta_2 = \frac{1}{21}\frac{a^\prime}{a}\left(PP^\prime-\frac{Q^\prime}{6}\right)-\frac{1}{18}\left(P+\frac{(P^\prime)^2}{2}\right).
\eeq
The function $Q(\eta)$ appearing in eq.~\eqref{eq:zeta_2} is the growing-mode solution of the equation
\beq
\label{eq:Q}
Q^{\prime\prime}+\frac{2a^\prime}{a}Q^\prime = \frac{5}{2}(P^\prime)^2-P.
\eeq
Note that the equality in eq.~\eqref{eq:phi2} is not exact: we have dropped additional terms in $\phi^{(2)}$ and $\psi^{(2)}$ which are negligible for the sub-horizon perturbations we will consider.

To calculate the predicted secondary temperature anisotropy pattern due to any isolated structure lying along the line-of-sight it only remains to specify the corresponding functional form of $F(\mathbf{x})$, as all the time-dependent terms in the equations above are uniquely determined by the choice of the background cosmology. We choose the form of $F$ so as to ensure that the first-order density perturbation $\delta$ today matches the spherically symmetric void profile specified in \cite{Finelli:2014}. This is achieved by inverting the Poisson equation, 
\beq
\label{eq:Poisson}
\nabla^2 F = -\frac{3\Omega_\rmn{m}H_0^2\delta}{\left(1-a^\prime P^\prime/a\right)|_{a=1}},
\eeq
to obtain  \cite{Nadathur:2011iu}
\bea
\label{eq:invPoisson}
 F(r) = &&\frac{3\Omega_\rmn{m}H_0^2}{\left(1-a^\prime P^\prime/a\right)|_{a=1}}\times \nonumber \\ 
&&\left[ \int_0^r\frac{r^{\prime2}}{r}\delta(r^\prime)dr^\prime + \int_r^\infty r^\prime\delta(r^\prime)dr^\prime\right],
\eea
which is generally valid for any spherically symmetric density perturbation $\delta(r)$. For the specific density profile given in eqn.~\eqref{eq:deltadef}, this means that $F(r) = F_0 e^{-r^2/r_0^2}$, where
\beq
\label{eq:F0}
F_0 = -\frac{\Omega_\rmn{m}H_0^2\delta_0 r_0^2}{2\left(1-a^\prime P^\prime/a\right)|_{a=1}}.
\eeq
Since at this order $\Phi= \phi^{(1)}=\psi^{(1)}$, our choice means that the gravitational potential has the same spatial dependence as  in \cite{Finelli:2014}, but we have also explicitly specified the time dependence.\footnote{In a later proceedings \cite{Szapudi:2014conf}, the authors of \cite{Finelli:2014} appear to claim a time dependence $\Phi\propto \eta^2$ for the potential, but this must be an error as it is not consistent with the standard time evolution in a $\Lambda$CDM background in eq.~\eqref{eq:phi1} above. Indeed such a time dependence would mean a potential fluctuation growing with time, which would result in an ISW temperature shift of the wrong sign.} 

\begin{figure}
\includegraphics[width=80mm]{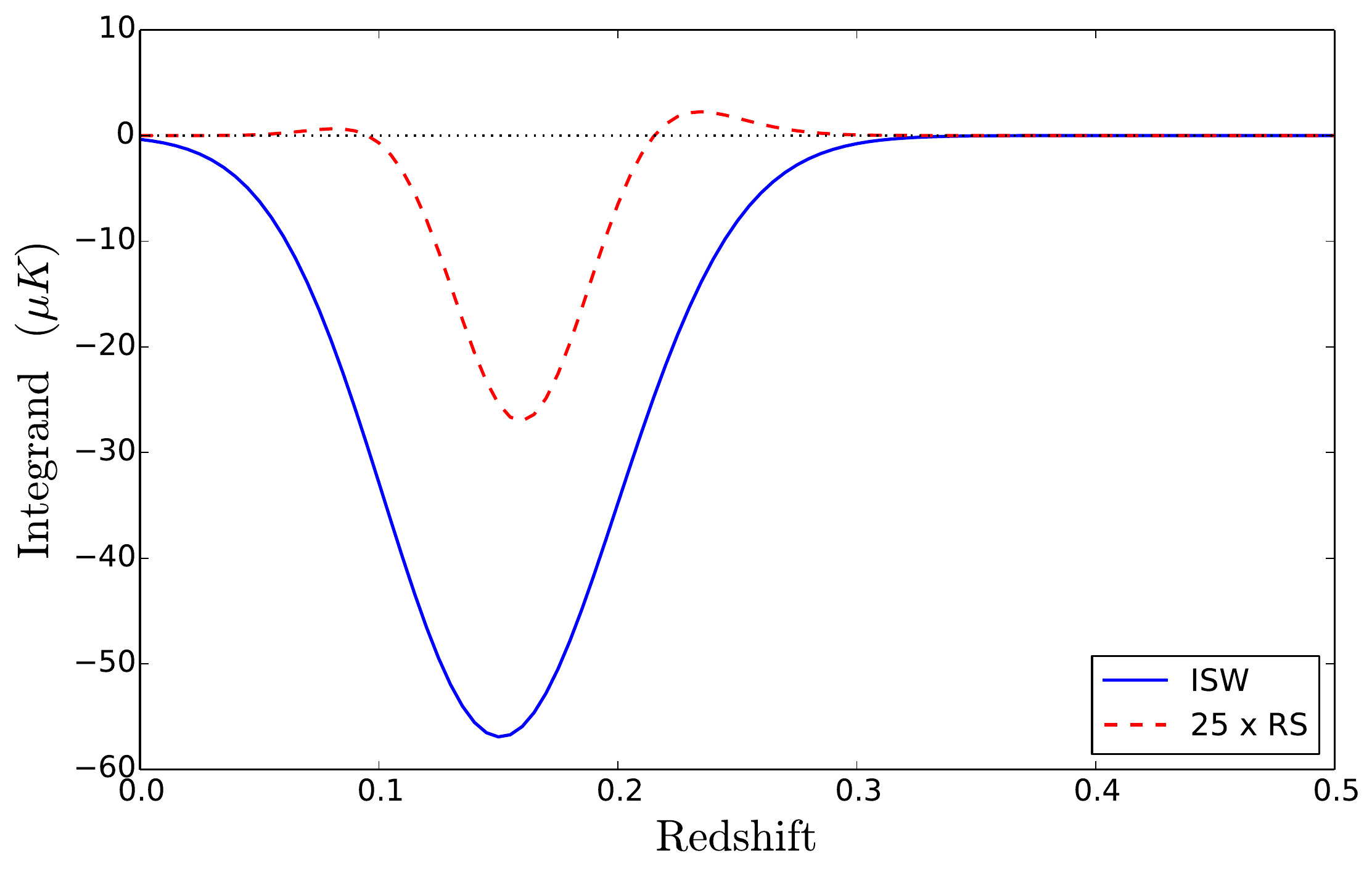}
\caption{The integrands appearing in the ISW and RS integrals in eqs.~\eqref{eq:deltaTISW} and \eqref{eq:deltaTRS} as a function of redshift $z$. The RS integrand has been multiplied by a factor of $25$ for clarity. The RS integrand is much smaller in magnitude and narrower in redshift than its ISW counterpart, and also changes sign at the overdense void edges.} 
\label{figure:integrands}
\end{figure}

Given this form of $F(r)$, the model of the void is specified by three numbers: the central underdensity $\delta_0$, the physical size $r_0$ and either the redshift of the centre of the void, $z_c$, or the comoving distance to the centre, $r_c$. One may use eqs.~\eqref{eq:I1}, \eqref{eq:phi1}, and \eqref{eq:P} to obtain the first-order ISW anisotropy as a function of the angle $\theta$ from the centre of the void:
\bea
\label{eq:deltaTISW}
\delta T_\rmn{ISW}(\theta) = \int_0^{z_\rmn{LS}} &&\left[ \left( \frac{a^{\prime\prime}}{a}-3\frac{a^{\prime2}}{a^2}\right)P^\prime + \frac{a^\prime}{a}\right] \times \nonumber \\
&&\frac{F_0 \exp\left(-\frac{\tilde{r}^2(z,\theta)}{r_0^2}\right)}{H(z)}dz,
\eea
where $\tilde{r}^2(z,\theta)=r^2(z)+r_c^2-2r(z)r_c \cos\theta$ with $r(z)$ the comoving distance to redshift $z$, $H(z)$ is the Hubble parameter at redshift $z$, and the integral in principle extends to the redshift of last scattering $z_\rmn{LS}$ though in practice it can be terminated much earlier for a sub-horizon sized void. Similarly, using eqs.~\eqref{eq:I2}, \eqref{eq:phi2} and \eqref{eq:Psi_0}, one obtains
\bea
\label{eq:deltaTRS}
\delta T_\rmn{RS}(\theta) = \int_0^{z_\rmn{LS}}\left( \frac{\tilde{r}^2(z,\theta)}{r_0^2}4\zeta_1^\prime+9\zeta_2^\prime\right) \times \nonumber \\
\frac{F_0^2 \exp\left(-\frac{2\tilde{r}^2(z,\theta)}{r_0^2}\right)}{r_0^2 H(z)}dz,
\eea
where $\zeta_1^\prime$ and $\zeta_2^\prime$ are complicated functions of $z$, whose full form is provided in Appendix \ref{appendixA}.

The void model that \cite{Finelli:2014} claim provides an RS temperature profile capable of fitting the Cold Spot has parameters $\delta_0=0.25$, $r0=195\;h^{-1}$Mpc and $z_c=0.155$ (or $r_c=450\;h^{-1}$Mpc), which we take as our fiducial model parameters. Figure~\ref{figure:integrands} shows the behaviour of the integrands of eqs.~\eqref{eq:deltaTISW} and \eqref{eq:deltaTRS} as functions of redshift $z$ along the line-of-sight passing through the centre of the void ($\theta=0$) for these fiducial parameters. Three important points are clear from this figure: the amplitude of the RS integral is much smaller than that of the linear ISW contribution; the RS integral receives contributions from a narrower range of redshifts, which also means that the RS effect contributes over a smaller angular range on the sky; and that at the void edges the RS contribution will be slightly positive due to the non-linear growth of structure in the void walls. This last point is also consistent with the discussion in \cite{Cai:2010hx}.

It is worth noting that eqs.~\eqref{eq:deltaTISW} and \eqref{eq:deltaTRS} do not admit simple closed-form analytic solutions in general ($\Lambda\neq0$), even for $\theta=0$, so it is not clear how the expression for $\delta T_\rmn{RS}(\theta)$ claimed in \cite{Finelli:2014} is derived. An Einstein-de Sitter universe ($\Omega_\rmn{m}=1$, $\Lambda=0$) provides a special case in which such an expression can be derived, as we discuss in Appendix \ref{appendixB}. Note that many previous studies of the RS effect of a void (e.g. \cite{Biswas:2008, Masina:2009a}) also assume an Einstein-de Sitter background; their results should therefore be compared with eqs. \eqref{eq:B4} and \eqref{eq:B5}.

\begin{figure*}
\includegraphics[width=160mm]{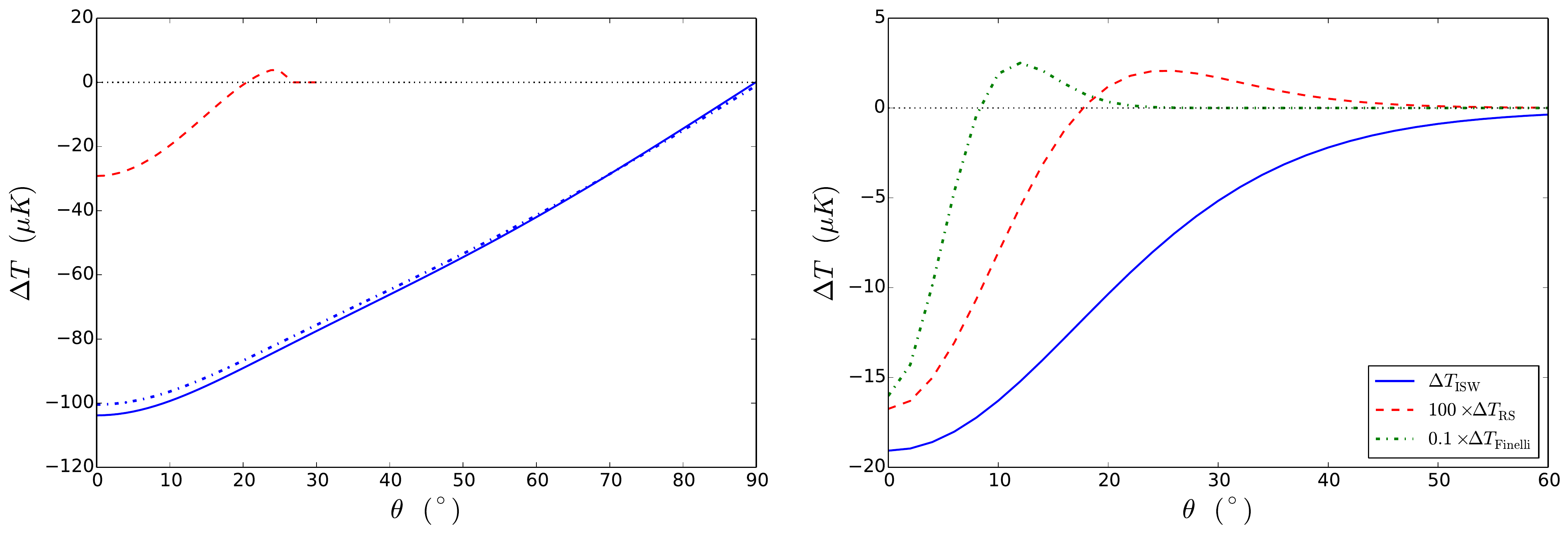}
\caption{\emph{Left panel}: The total temperature anisotropy $\Delta T(\theta)$ due to the fiducial void profile with $\delta_0=0.25$, $r_0=195\;h^{-1}$Mpc and $z_c=0.155$, as calculated with the LTB model (blue solid line), and with perturbation theory (blue dot-dashed line). In both cases, the dominant contribution is a dipole term. The red dashed curve shows $\Delta T(\theta)$ for the cLTB model, which does not show a dipole. \emph{Right panel}: The angular dependence of $\Delta T_\rmn{ISW}$ (blue solid curve) and $\Delta T_\rmn{RS}$ (red dashed) for the fiducial void, as calculated from eqs.~\eqref{eq:deltaTISW} and \eqref{eq:deltaTRS}. The RS anisotropy is magnified by a factor of 100 for clarity. The green (dot-dashed) line shows the $\Delta T(\theta)$ claimed in \cite{Finelli:2014}, multiplied by 0.1 for clarity.} 
\label{figure:Tprofiles}
\end{figure*}

\subsection{Results}
\label{section:theoryresults}

Figure~\ref{figure:Tprofiles} shows the temperature anisotropy profiles $\Delta T(\theta)$ for the void defined by eq.~\eqref{eq:deltadef}, with the fiducial parameters given above, as calculated using both the exact LTB and perturbation theory approaches. We also show $\Delta T(\theta)$ for the corresponding cLTB model.

The first thing that is obvious from this figure is that the results from the perturbation theory and LTB approaches match each other quite well, but both differ markedly from the claimed RS temperature profile for the same void parameters in \cite{Finelli:2014}. The amplitude of the signal we find is approximately a factor of a third smaller, but is at least of the same order of magnitude, $\mathcal{O}(100\;\upmu\rmn{K})$. However, the angular scale of the profile we find is much broader, extending out to $\theta=90^\circ$.

To understand this, let us break up the total temperature anisotropy obtained in Section~\ref{section:2PT} into pieces corresponding to the monopole, dipole, first-order ISW and second-order RS effects. Comparison of their relative amplitudes shows that $\Delta T_\rmn{mon}=-0.95\;\upmu$K, $\Delta T_\rmn{dip}|_{\theta=0}=-80.3\;\upmu$K, $\Delta T_\rmn{ISW}|_{\theta=0}=-19.1\;\upmu$K and $\Delta T_\rmn{RS}|_{\theta=0}=-0.17\;\upmu$K, where we have taken $T_\rmn{CMB}=2.7255$K \cite{Fixsen:2009}. That is, for the void parameters claimed in \cite{Finelli:2014}, the dominant temperature anisotropy is in fact a \emph{dipole} term, and the RS contribution is, as expected based on the arguments in Section~\ref{section:heuristics}, negligible compared to the linear ISW integral. In the right panel of Figure~\ref{figure:Tprofiles} we show the angular dependence of the ISW and RS terms, with the claimed profile from \cite{Finelli:2014} included for comparison.

The origin of the dipole term itself is worth further consideration. The void model considered here corresponds to a density profile that is only asymptotically compensated at infinity, as seen from eq.~\eqref{eq:deltadef}, so the gravitational potential $\Phi$ only approaches zero at $r\rightarrow\infty$. In particular, since for these void parameters  $r_c  \simeq 2.3 r_0$, $\Phi$ is not negligible at the observer location. In other words, the void is so large and so close that the observer lies within its potential and is moving appreciably with respect to its centre. On the other hand, for the compensated profile of the cLTB model, $\Phi$ goes to zero at the compensation radius $r_b$ so there is no dipole contribution to the temperature profile.

The magnitude of the dipole produced by this void is much smaller than the total kinetic dipole seen by Planck \cite{Planck:EppurSiMuove}, and it will not be visible in dipole-subtracted CMB maps. All that will be left is $\Delta T_\rmn{ISW}(\theta)+\Delta T_\rmn{RS}(\theta)$, which is far too small to explain the observed Cold Spot temperature decrement of $\sim150\;\upmu$K, and in any case has the wrong angular profile.

\begin{figure}
\includegraphics[width=85mm]{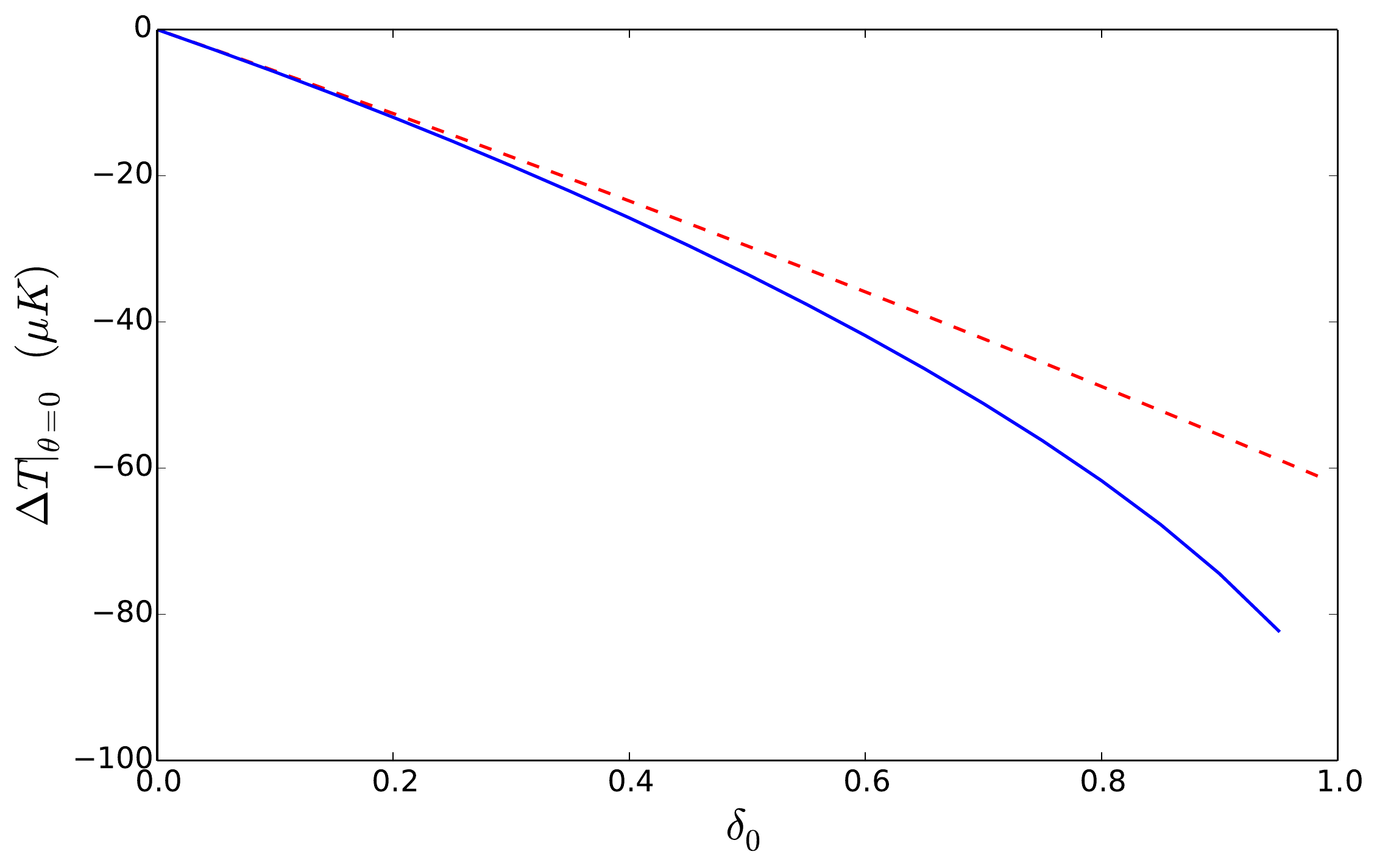}
\caption{The dependence of the total temperature anisotropy $\Delta T_\rmn{ISW}+\Delta T_\rmn{RS}$ for the fiducial void profile, as a function of $\delta_0$ for a void of size $r_0=195\;h^{-1}$Mpc, as calculated using the full LTB approach (blue solid line) and the perturbation theory approach (red dashed line). The redshift of the void centre is fixed at $z_c=0.4$ so that the dipole contribution to the LTB calculation is negligible.} 
\label{fig:d0scaling}
\end{figure}

To study the applicability of the perturbation theory approach to calculating ISW temperature anisotropies in more detail, we use the fiducial density profile of eq.~\eqref{eq:deltadef} and fix the void radius to be $r_0=195\;h^{-1}$Mpc as before. However, we place the centre of the void at $z_c=0.4$, far enough out that the monopole and dipole contributions to $\delta T^{(1)}$ are negligible. Figure~\ref{fig:d0scaling} then shows the scaling of $\Delta T|_{\theta=0}$ versus the central underdensity $\delta_0$ as calculated using the exact LTB model and perturbation theory to second order. Whereas the two results agree very well for small $\delta_0$ values, for $\delta_0\gtrsim0.5$ the perturbation theory approach is unable to match the fully non-linear LTB result. The importance of such non-linear evolution for LTB models has been noted before in other contexts \cite{Bolejko:2013}.

\subsection{Estimating void probabilities}
\label{section:voidprobs}

Although it is clear from the results above that the specific void claimed to have been detected by \cite{Szapudi:2014} cannot possibly explain the Cold Spot, we are interested in the more general question of whether \emph{any} reasonable void can do so. To answer this question we shall restrict ourselves to the case of the fiducial density profile and perform a scan over parameters $\delta_0$ and $r_0$, using the full LTB approach to calculate the temperature shift $\Delta T|_{\theta=0}$ only.\footnote{The shape of $\Delta T(\theta)$ for the fiducial profile will actually never be able to match the observed Cold Spot temperature profile, since it does not cross zero at any angle. For this a compensated void such as the cLTB model is required. However, as a first approximation we need only consider the magnitude of the temperature decrement at the centre, and for the same choices of $\delta_0$ and $r_0$ voids with compensated profiles such as cLTB produce similar values of $\Delta T|_{\theta=0}$ but are far less likely in $\Lambda$CDM.} In order to ensure that the unwanted dipole does not contribute to this value, we require that the distance to the centre $r_c$ be large enough. This is done by setting the ratio $r_0/r_c$ to be constant, such that the void subtends an angle of $\sim10^\circ$ on the sky.

For each choice of $(\delta_0,r_0)$ we also wish to evaluate the likelihood of existence of such a void in a $\Lambda$CDM universe. To do this we calculate the Gaussian-filtered density contrast at the centre of the void, when the filter width is set equal to $r_0$, in units of the rms density fluctuation $\sigma(r_0)$ at the same scale and using the same Gaussian filter.\footnote{The use of a Gaussian filter rather than a top-hat is necessary to ensure the convergence of higher-order moments of the density field used in the number density calculation \cite{BBKS}.} We then use standard peaks theory for a Gaussian random field \cite{BBKS} to estimate the cumulative number density of peaks of the matter density field which represent an equal or greater negative fluctuation, quantified by $\nu=\delta_\rmn{filt}/\sigma_0$. By multiplying this number density by the total volume of the universe enclosed within redshift 0.5, we obtain an estimate of the number of voids of equal or greater ``extremeness" that we should expect to observe within the local universe in a $\Lambda$CDM cosmology. 

\begin{figure}
\includegraphics[width=85mm]{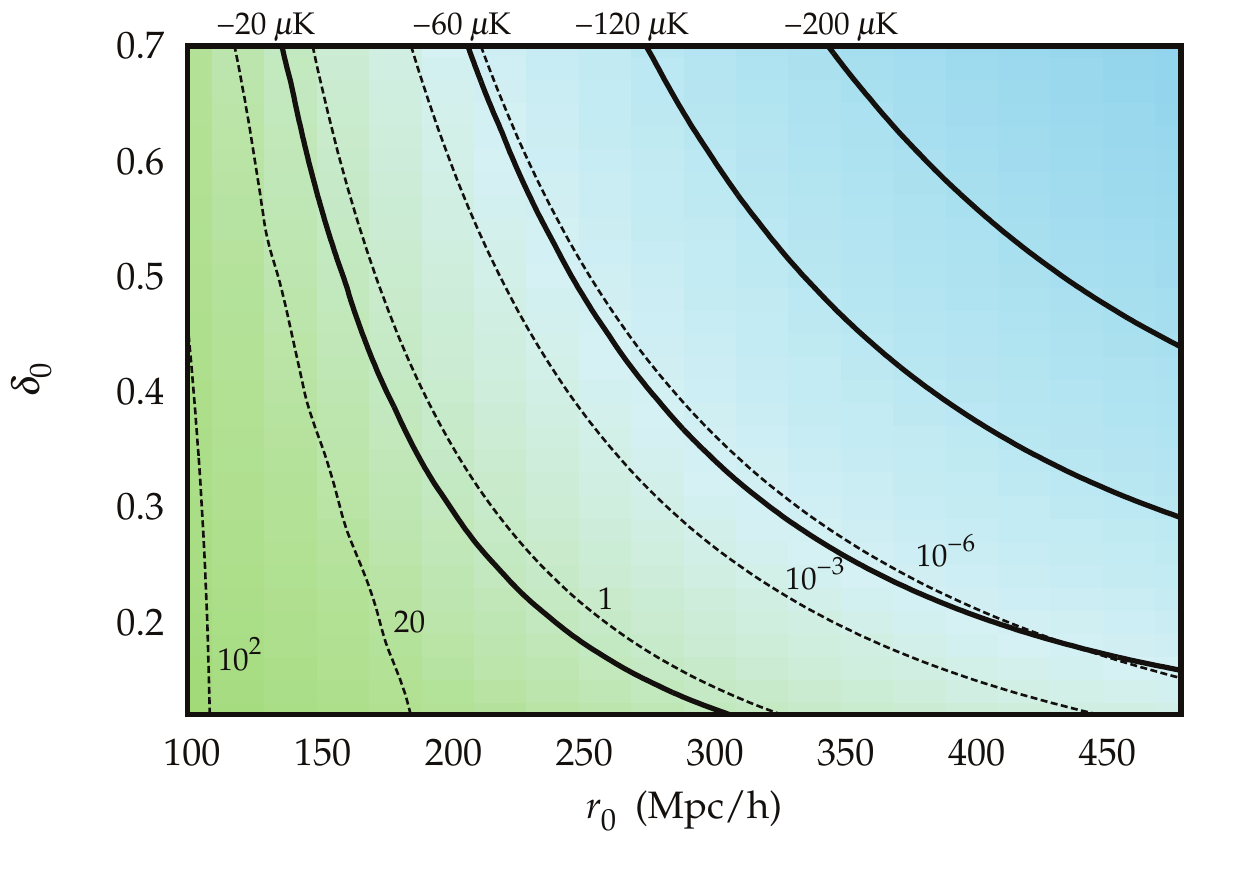}
\caption{The colour scale shows the dependence of $\Delta T|_{\theta=0}$ on void parameters $\delta_0$ and $r_0$ calculated with the LTB model. The distance to the void centre is adjusted such that voids of different sizes subtend the same angle on the sky, and the dipole effect is zero. Solid lines are contours of equal $\Delta T|_{\theta=0}$, at values $-20\;\upmu$K, $-60\;\upmu$K, $-120\;\upmu$K, and $-200\;\upmu$K (labelled on top). The dashed lines are contours of equal ``extremeness", labelled by the approximate number of equally extreme voids expected to exist within $z<0.5$ in a $\Lambda$CDM universe. Where these numbers are $<1$, they may be taken as the probability of existence of a single such void.} 
\label{fig:contours}
\end{figure}

In Figure~\ref{fig:contours} we show the results of these two calculations: the solid lines show contours of equal $\Delta T|_{\theta=0}$, and the dashed lines contours of equal expected number of voids. When the expected number values are $<1$, they can be regarded as representing the probability of finding a single void of such extremeness in $\Lambda$CDM. 

Note that for the likelihood calculation we use the fiducial density profile of eq.~\eqref{eq:deltadef} to calculate $\delta_\rmn{filt}$; for a compensated top-hat profile like the cLTB, $\delta_\rmn{filt}$ will be much larger and therefore the void correspondingly less likely. It is also worth mentioning that choosing the filter radius to be equal to $r_0$ is a somewhat arbitrary choice. In fact the normalized void density fluctuation $|\nu|$ peaks at smaller filter radii $\sim0.5r_0$, so our choice is somewhat conservative and does not minimize the void likelihood. On the other hand, our treatment is in any case approximate and a proper calculation of the likelihood is beyond the scope of this paper. The likelihood contours in Figure~\ref{fig:contours} should be treated as qualitative guides rather than precise values.

However, some important general conclusions can be drawn. Voids such as that postulated in \cite{Finelli:2014}, with $\delta_0=0.25$ and $r_0=195\;h^{-1}$Mpc, are not particularly unlikely in a $\Lambda$CDM universe -- we should expect to see $\sim10-20$ of them, as argued in Section~\ref{section:heuristics}. Such voids also produce a temperature effect that is far too small to explain the Cold Spot. On the other hand, the probability of existence of voids drops off much more rapidly than the possible temperature signal increases, such that voids capable of producing $\Delta T|_{\theta=0}=-60\;\upmu$K are already exceedingly unlikely and the probability that any void could explain the full Cold Spot temperature decrement of $-150\;\upmu$K is very small within the $\Lambda$CDM model. This conclusion is in agreement with the earlier results \cite{Cai:2010hx,Nadathur:2014a,Hotchkiss:2014,Nadathur:2014b} discussed in Section~\ref{section:heuristics}.

It is also worth noting that for combinations $(\delta_0,r_0)$ for which voids are relatively likely to exist, not only is the $\Delta T$ calculated with the LTB model small in absolute magnitude, but the result is also well approximated by perturbation theory.

So far we have made the unrealistic assumption that a single void along the line of sight contributes all the temperature anisotropy of the Cold Spot, without considering the role of intrinsic fluctuations on the last scattering surface. We turn to this in the next Section.



\section{Comparing the supervoid hypothesis to data}
\label{section:Bayes}

Given any model of a hypothesized supervoid along the line of sight and the temperature profile $\delta T(\theta)$ due to it, we would like both to determine how good a fit to the Cold Spot data this model provides, and to compare this goodness of fit to appropriate alternative explanations for the Cold Spot. In \cite{Finelli:2014} the authors attempt to address this question by constructing a $\chi^2$ statistic for the fit to the Cold Spot temperature profile as
\beq
\label{eq:chi2}
\chi^2 = \sum_{ij}\left(\delta T^\rmn{th}_i-\delta T^\rmn{CMB}_i\right)C^{-1}_{ij}\left(\delta T^\rmn{th}_j-\delta T^\rmn{CMB}_j\right),
\eeq
where $\delta T^\rmn{CMB}_i$ represents the observed average CMB temperature in angluar bins centred at angles $\theta_i$ from the centre of the Cold Spot, $\delta T^\rmn{th}_i$ is the theoretical prediction for the temperature at $\theta_i$, $C_{ij}$ is the covariance matrix of the CMB determined from simulated random maps, and the sums run over all bins.

As shown in the previous section, the estimate of the magnitude of the RS effect of a void provided in \cite{Finelli:2014} is three orders of magnitude too large, rendering the $\chi^2$ values they calculate using this predicted profile meaningless. If one wants to test the hypothesis that the Cold Spot is attributable \emph{in its entirety} to a supervoid, one should use $\delta T^\rmn{th} = \delta T_\rmn{ISW} + \delta T_\rmn{RS}$ calculated from eqs.~\eqref{eq:deltaTISW} and \eqref{eq:deltaTRS}; for the void parameters claimed in \cite{Finelli:2014} -- and indeed for any reasonable void parameters -- this hypothesis would instead prove a very poor fit to the data.

But there is a more subtle and general problem with using the $\chi^2$ value obtained from \eqref{eq:chi2} for model comparisons, and that is the choice of the null hypothesis with which to compare the supervoid hypothesis. Ref. \cite{Finelli:2014} assumes as the null hypothesis the model with prediction $\delta T^\rmn{th}(\theta)=0,$\footnote{J. Garcia-Bellido, private communication.} which amounts to assuming that the Cold Spot is a randomly chosen point on the CMB.  This is obviously not the case: the Cold Spot direction is special by construction, having been specifically selected because when the CMB map is filtered using a spherical Mexican hat wavelet (SMHW) it is the coldest direction on the sky. This constitutes an enormous \emph{a posteriori} selection effect.

\begin{figure}
\includegraphics[width=85mm]{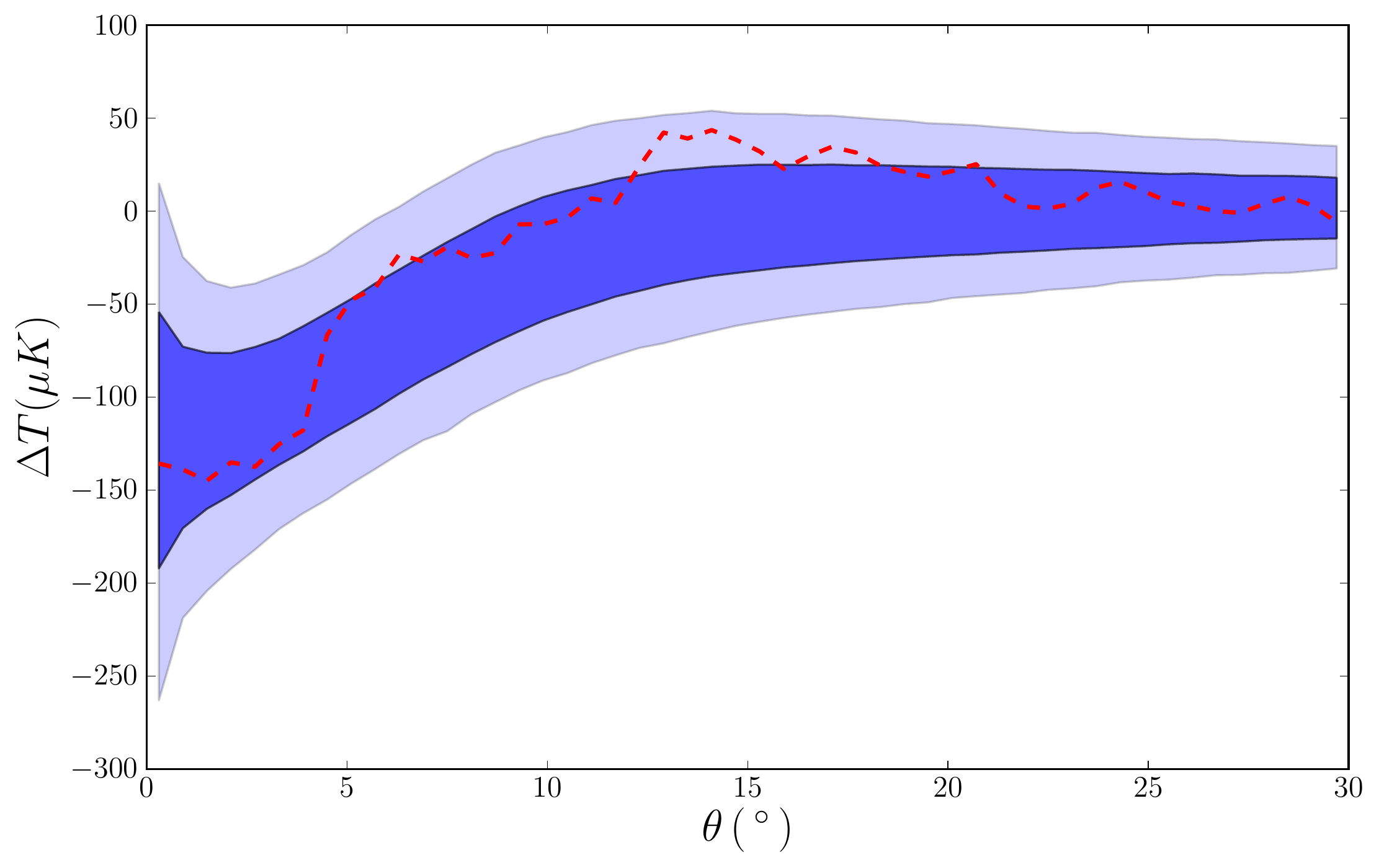}
\caption{The average temperature profile around the Cold Spot direction (red dashed line) measured from the Planck SMICA map, showing the central cold region and the surrounding hot ring at $\theta\sim15^\circ$. The shaded regions show the 68\% and 95\% C.L. range in the temperature profiles around the coldest spots identified in 10,000 random Gaussian maps using the same SMHW technique.}
\label{fig:CStempbands}
\end{figure}

To ensure a fair comparison between models the null hypothesis must correctly account for this selection effect on the profile. That is to say, if one wishes to test the null hypothesis that the observed temperature profile around the direction of the Cold Spot is simply due to a statistical fluctuation, one must at the very least compare the observed temperature profile with the average profile found around the coldest spot obtained when random CMB maps are filtered using the same SMHW filter with the same angular width. Indeed one may also wish to investigate the possibility of further \emph{a posteriori} effects in the choice of filter and angular width \cite{Zhang:2010}, but we will restrict ourselves to the minimal case in this paper.

To demonstrate the importance of this selection effect, we perform the following analysis. First we generate 10,000 random synthetic Gaussian CMB maps with the help of the {\small HEALPix} software package \cite{Gorski:2004by} using the $C_\ell$ values for the best-fit Planck power spectrum \cite{Planck:CosmoParams}, each at $N_{side}=128$. To each of these random maps we apply the Planck union (U74) sky mask, also downgraded to $N_{side}=128$. We then filter each map using an SMHW of width $6^\circ$, which roughly corresponds to the filter width giving the maximum significance for the real Planck/WMAP Cold Spot. We then select the pixel direction, from the set of pixels on a {\small HEALPix}, $N_{side}=16$ map, that corresponds to the coldest filtered temperature. To match past convention, we wish to ensure that the significance of the filtered signal in the chosen pixel is not dominated by mask effects. To do this, we also filter the ($N_{side}=128$) mask with the square of the SMHW filter and ignore any ($N_{side}=16$) pixel direction in the filtered mask with a value $<0.95$.\footnote{We use the square of the SMHW filter because it changes sign. This means that even regions where the mask is uniformly set to $1$ could give a filtered signal $<1$ unless using the squared filter.}. Also, note that when we downgrade the U74 mask this results in some pixels being partially masked. We also ignore any pixel direction with a value $<0.9$ in the unfiltered, downgraded mask. We then measure the average temperature profile in concentric rings about these coldest spots as a function of the angular distance $\theta$ from the central direction in each (unfiltered, $N_{side}=128$) map. Finally we repeat exactly the same procedure for the Planck SMICA map \cite{Planck:Overview}, recovering the actual temperature profile around the Cold Spot and its SMHW-filtered temperature.

The mean of these profiles from random maps provides an estimate of the effect of the selection effect on the observed Cold Spot temperature. In Figure~\ref{fig:CStempbands} we show the actual Cold Spot profile overlaid on top of the $68\%$ and $95\%$ C.L. (more precisely, the equivalent Gaussian 1- and 2-$\sigma$) bands determined from the random maps. This is the correct null hypothesis to which the supervoid hypothesis (or indeed any other proposed exotic explanation) must be compared. The first and most important conclusion to be drawn from this figure is that the mere fact that the Cold Spot was chosen precisely because it was cold \emph{alone} can satisfactorily account for the temperature decrement at the centre of the Cold Spot.

Nevertheless, in agreement with the Planck analysis \cite{Planck:Isotropy}, we find that the real Cold Spot is still unusual at the $\sim3\sigma$ level in $\Lambda$CDM, in that fewer than $1\%$ of the coldest spots in random maps have as cold a total filtered temperature. The reason for this lies in the fact that, as also explained in \cite{Zhang:2010}, the true Cold Spot temperature profile shows a transition from a cold centre to a hot surrounding ring at $\theta\simeq15^\circ$ which happens to roughly coincide with the change in sign of the SMHW profile. In other words, despite the fact that the Cold Spot profile does not at any point lie outside the $95.5\%$ C.L. bands in Figure~\ref{fig:CStempbands}, what is unusual is that it crosses from being slightly colder than average at $\theta\simeq0^\circ$ to being slightly hotter than average at $\theta\simeq15^\circ$.

\begin{figure}
\includegraphics[width=85mm]{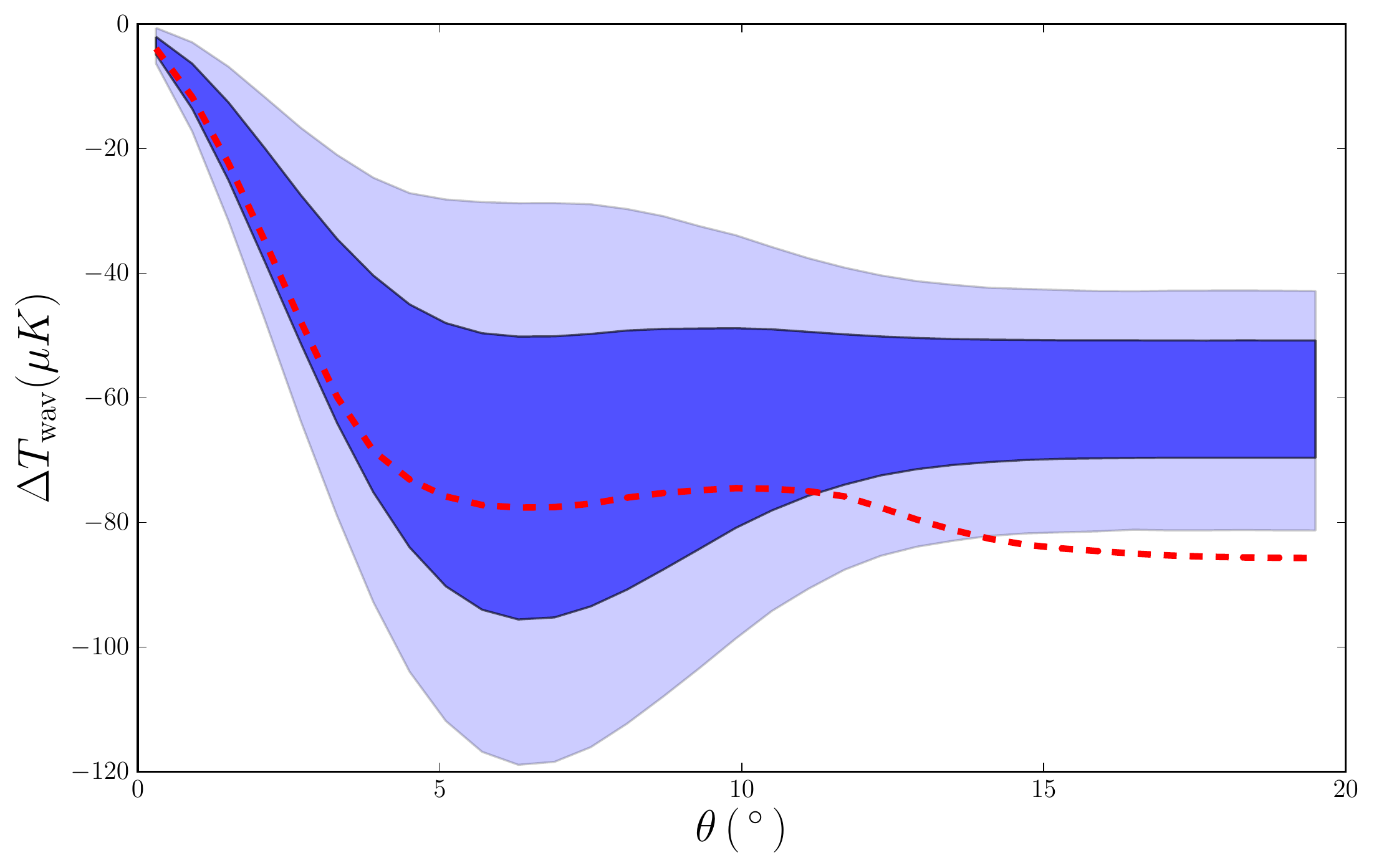}
\caption{The red dashed line shows the cumulative filtered temperature $\Delta T_\rmn{wav}(\theta)$ (eq.~\eqref{eq:cumulwavelet}) using an SMHW of width $6^\circ$ for the Cold Spot in the Planck SMICA map. The shaded regions indicate the $68\%$ and $95\%$ confidence limits determined from applying the same procedure to the coldest spots in 10,000 random CMB maps. The filtered temperature for the real Cold Spot only becomes statistically significant at $\theta\sim15^\circ$.}
\label{fig:whereisanomaly}
\end{figure}

To demonstrate this another way, we calculate the cumulative filtered signal out to angle $\theta$ as
\beq
\label{eq:cumulwavelet}
\Delta T_\rmn{wav}(\theta,R) = \int_{0}^\theta \int_{0}^{2\pi} \Delta T(\theta^\prime,\phi^\prime)\Psi(\theta^\prime,\phi^\prime;R)d\Omega^\prime,
\eeq
where $\Psi(\theta^\prime,\phi^\prime;R)$ is the value of the SMHW of width $R$ at $(\theta^\prime,\phi^\prime)$. In Figure~\ref{fig:whereisanomaly} we show this cumulative signal as a function of $\theta$ for the Planck SMICA map, compared to the $95\%$ C.L. band obtained from the corresponding values for the coldest spots in random maps. Our Cold Spot first becomes anomalous at angles $\theta\sim15^\circ$, showing that it is not the central cold region that is anomalous, but the particular combination of the cold region and the hot surrounding ring.
 
From Figures~\ref{fig:contours} and \ref{fig:CStempbands} we conclude that the temperature contribution to the Cold Spot of any supervoid is dwarfed by the selection effect inherent in identifying the Cold Spot direction. Nevertheless, a void aligned along the line of sight could still contribute \emph{part} of the temperature effect seen, and perhaps this could help to explain the residual anomaly discussed above.

At first sight, it may appear that evaluating the likelihood of such a scenario has become more complicated, since one must now estimate the expected temperature profile due to the coldest fluctuation on the last scattering surface (in the absence of a void), estimate the additional contribution that could arise from the most extreme supervoid that one could expect to be present in a $\Lambda$CDM universe, and also estimate the probability that this supervoid is by chance aligned such as to enhance the effect of the rare fluctuation on the last scattering surface. 

However, we find that for combinations of void parameters that are likely to occur within $\Lambda$CDM, the standard linear ISW treatment represents a good approximation to the full non-linear treatment of the resulting temperature anisotropy using the LTB model. The $C_\ell$ values used to generate the random maps used in producing Figures~\ref{fig:CStempbands} and \ref{fig:whereisanomaly} already include this secondary linear ISW contribution.\footnote{In fact, removing this contribution by hand hardly affects the results, demonstrating the miniscule probability of having a suitable supervoid aligned by chance with the direction of the appropriate fluctuation on the last scattering surface.} Therefore comparing the coldest spots in  random CMB maps to our own Cold Spot as is usually done \emph{already accounts for the contribution of any possible supervoid along the line of sight}. Therefore, within $\Lambda$CDM, insofar as the Cold Spot is anomalous, the additional hypothesis of a supervoid cannot help resolve this anomaly.


\section{Discussion}
\label{section:discussion}

In this paper we have explored the possibility that the Cold Spot seen in the CMB is the result of a large void along the line of sight. In particular we have examined in detail the claim in \cite{Finelli:2014} that a recently reported void found in the WISE-2MASS galaxy data \cite{Szapudi:2014}, is capable of accounting for the Cold Spot temperature profile through the second-order Rees-Sciama effect. We find this claim to be mistaken. We have calculated the true temperature effect of the postulated void in two different ways, using an exact LTB solution and perturbation theory. We find that the second-order Rees-Sciama effect is much smaller than the linear order ISW anisotropy. The total temperature effect of such a void is in fact dominated by a dipole due to the observer's motion with respect to its centre, but this contribution is small enough to be unobservable in dipole-subtracted CMB maps. When the dipole effect is subtracted, the remaining ISW contribution is far too small to account for the Cold Spot, and in any case produces a temperature profile of the wrong shape. We have further shown that in order to produce $\Delta T\sim-150\;\upmu$K as seen at the Cold Spot location a void would need to be so large and so empty that within the standard $\Lambda$CDM framework the probability of its existence is essentially zero.

Note that we have not at any point questioned the claim in \cite{Szapudi:2014} that a large void actually exists at this location in the direction of the Cold Spot. Indeed we have argued that the size and density contrast values reported for this void are not extraordinary, and that one should expect to find several such voids within the reasonably local universe (i.e. at redshifts $<0.5$). Some earlier examples of such voids have already been reported \cite{Nadathur:2014a}. Instead our argument is based on the fact that even if such a void does exist, the maximum temperature effect it could produce on the CMB is still insufficient to explain the Cold Spot.

In fact our results provide a more general argument against any supervoid explanation of the Cold Spot. This is because we find that for density contrasts and sizes of voids that are compatible with their existence in a $\Lambda$CDM universe, the linear theory calculation of the induced ISW temperature shifts is sufficient, and that these temperature shifts are small. Therefore the standard estimation of the statistical significance of the Cold Spot anomaly, which makes use of the CMB power spectrum in a $\Lambda$CDM model that includes a linear ISW contribution, already self-consistently incorporates the possible contribution of any void along the line of sight within the $\Lambda$CDM framework. This means the existence of any structure capable of explaining the Cold Spot  must be at least as anomalous in $\Lambda$CDM as the existence of the Cold Spot itself. 

Of course the $\Lambda$CDM model may be incomplete, and rare supervoids at odds with its predictions may in principle exist, so it is worth searching for them in observational data. However, if such a supervoid did exist, it would point to a significant failure of our theories of structure formation or initial conditions, which have been otherwise successful in fitting a variety of cosmological data. Therefore, unless and until such a structure is actually found, or a good theoretical motivation for it is provided, the sensible choice of priors must reflect a presumption against its existence. Any comparison of the Bayesian evidence for a hypothetical supervoid model must then also account for these priors. We argue that doing so correctly will disfavour the supervoid over random statistical fluctuations on the last scattering surface as an explanation for the Cold Spot anomaly.

In the process of making this argument, we have also shown that it is actually not the coldness of the temperature fluctuation at the centre of the Cold Spot which is anomalous. The mere fact that the Cold Spot was specifically selected as the coldest spot in our CMB (after applying an SMHW filter) is already a sufficient explanation of this: Figure \ref{fig:CStempbands} shows that at its centre our Cold Spot is well within the expected range of temperatures of the coldest spots on random CMB maps. Instead it is the \emph{combination} of the cold centre and the hot ring at larger angles which makes our Cold Spot unusual. This was perhaps not entirely unknown -- for instance, it is the reason that the Cold Spot looks anomalous when using an SMHW  but not under other filters  \cite{Zhang:2010} -- but it provides an alternative interesting perspective to the problem. 

To achieve such cold centres and hot rings through the ISW effect requires voids with the extreme compensated top-hat type of density profile shown in the ``cLTB" model we consider. This is the theoretically motivated end-state profile for highly evolved and non-linear voids on very small scales \cite{Sheth:2003py}; however for voids on $\gtrsim100\;h^{-1}$Mpc scales such a density distribution is exceedingly unlikely, thus further disfavouring a supervoid explanation.

\emph{Note added}: After preparation of this manuscript, we became aware of another preprint \cite{Zibin:2014note}, which appeared on the arXiv almost simultaneously and independently examines some of these same issues. Despite small differences in approach, including the choice of gauge and the mapping between LTB and perturbed FRW models, in the region of overlap the broad conclusions of that work -- that the dominant contribution of the void proposed by \cite{Szapudi:2014} is a dipole anisotropy, the second-order RS term is subdominant to the ISW term, and the total ISW+RS effect is much smaller than claimed by \cite{Finelli:2014} -- closely match those presented here.

\begin{acknowledgments}
We thank Juan Garcia-Bellido and Francesco Paci for correspondence and discussions. SN acknowledges support from Academy of Finland grant 1263714. SH received research funding from the European Research Council under the European Union’s Seventh Framework Programme (FP/2007–2013) / ERC Grant Agreement No. [308082].
\end{acknowledgments}

\appendix

\section{Time-dependent terms in $\delta T_\rmn{ISW}$ and $\delta T_\rmn{RS}$ }
\label{appendixA}

The functions $\zeta_1^\prime$ and $\zeta_2^\prime$ appearing in eq.~\eqref{eq:deltaTRS} can be written as \cite{TomitaInoue:2008}
\beq
\zeta_1^\prime = \frac{1}{4}\left[P^\prime-\frac{a^\prime}{a}\left(P+P^{\prime2}\right)+\left(3\frac{a^{\prime2}}{a^2}-\frac{a^{\prime\prime}}{a}\right)PP^\prime\right],
\eeq
and
\bea
\zeta_2^\prime = -\frac{1}{9}P^\prime &+& \frac{1}{18}\frac{a^\prime}{a}P + \frac{5}{36}\frac{a^\prime}{a}P^{\prime2} \nonumber \\
&+&\frac{1}{21}\left( \frac{a^{\prime\prime}}{a}-3\frac{a^{\prime2}}{a^2}\right)\left(PP^\prime-\frac{Q^\prime}{6} \right),
\eea
where we have used eq~\eqref{eq:Q} and the fact that 
\beq
P^{\prime\prime}+2\left(\frac{a^\prime}{a}\right)P^\prime-1=0.
\eeq
Some additional useful relations are 
\beq
\frac{a^\prime}{a}=aH =aH_0\sqrt{\Omega_\rmn{m}a^{-3}+\Omega_\Lambda} ,
\eeq
and
\beq
\frac{a^{\prime\prime}}{a}=\frac{H_0^2}{2a}\left(\Omega_\rmn{m}+4\Omega_\Lambda a^3\right).
\eeq

\section{An $\Omega_\rmn{m}=1$ background }
\label{appendixB}

A background Einstein-de Sitter cosmology ($\Omega_\rmn{m}=1$, $\Lambda=0$) represents a special case in which the general equations presented above simplify appreciably \cite{Mollerach:1997}. Firstly, we have
\beq
P^\prime = \frac{2}{5aH},
\eeq
so that the first-order potential terms $\phi^{(1)}$ and $\psi^{(1)}$ are time-independent and $\delta T_\rmn{ISW}$ vanishes, as expected. In addition, we obtain $\zeta_1 = \frac{3}{200}\eta^2$ and $\zeta_2 = -\frac{1}{210}\eta^2$, so that for our fiducial profile eq.~\eqref{eq:deltaTRS} reduces to:
\beq
\delta T_\rmn{RS} = \int_0^{\eta_0} \frac{3\eta}{25}\left( \frac{\tilde{r}^2}{r_0^2} - \frac{5}{7}\right)\frac{F_0^2}{r_0^2}e^{-\frac{2\tilde{r}^2}{r_0^2}}d\eta,
\eeq
where
\beq
\tilde{r}^2(\theta,\eta) = r^2 +r_c^2-2rr_c\cos\theta
\eeq
as before, and $r(\eta)=\eta_0-\eta=2/H_0-\eta$.

Even in this simplified case, the full $\theta$-dependent integral is very tedious and is better evaluated numerically. However, by making some judicious approximations regarding the ratios $\eta_0/r_0$ and $r_c/r_0$, we can obtain a simple expression for the maximum amplitude of the RS signal through the line passing through the void centre,
\beq
\label{eq:B4}
\delta T_\rmn{RS}|_{\theta=0} = -\frac{13}{168}\sqrt{\frac{\pi}{2}}\delta_0^2\left(r_0H_0\right)^3\left(1-\frac{r_cH_0}{2}\right).
\eeq
For the fiducial void parameters $\delta_0=0.25$, $r_0=195\;h^{-1}$Mpc, $z_c=0.155$, one can check that this leads to an amplitude of the RS signal
\beq
\label{eq:B5}
\Delta T_\rmn{RS}|_{\theta=0}\simeq-4.2\;\upmu\rmn{K},
\eeq
which is still orders of magnitude smaller than the value claimed in \cite{Finelli:2014}. This value is in good agreement with previous results which also assume $\Omega_\rmn{m}=1$, e.g. \cite{Masina:2009a}. Also note that the magnitude of the RS term is maximum in the case $\Omega_\rmn{m}=1$ and will be smaller than this for any model with non-zero $\Lambda$ \cite{TomitaInoue:2008}.

\bibliography{refs.bib}
\bibliographystyle{apsrev.bst}

\end{document}